\documentclass[10pt,draft,onecolumn]{IEEEtran}
\IEEEoverridecommandlockouts
\usepackage[final]{graphicx}
\usepackage{srcltx}
\usepackage{cite}

\usepackage{amssymb}
\usepackage{amsmath}   
\usepackage{amsthm}
\usepackage{bm}
\usepackage{subcaption}
\usepackage{xcolor}
\usepackage{balance}
\usepackage{float}

\begin{document}

\title{On Low Complexity Detection for QAM Isomorphic Constellations}

\author{Farbod Kayhan \\
Interdisciplinary Centre for Security, Reliability and Trust (SnT),\\ University of Luxembourg \\(email: farbod.kayhan@uni.lu). 
}

\newtheorem{Theorem}{Theorem}[section]
\newtheorem{Lemma}{Lemma}[section]
\newtheorem{Definition}{Definition}[section]
\newtheorem{Conjecture}{Conjecture}[section]
\newtheorem{Corollary}{Corollary}[section]
\newtheorem{Proposition}{Proposition}[section]

\newcommand{\C}{\mathcal{C}}
\newcommand{\SPSK}{\text{-S}\text{PSK}}
\newcommand{\SQPSK}{\text{-S}\text{QPSK}}
\newcommand{\SAPSK}{\text{-S}\text{APSK}}
\newcommand{\SkPSK}{\text{-S}^k\text{PSK}}
\newcommand{\SkQPSK}{\text{-S}^k\text{QPSK}}
\newcommand{\SkAPSK}{\text{-S}^k\text{APSK}}
\newcommand{\PSNR}{\text{PSNR}}
\newcommand{\SNR}{\text{SNR}}
\newcommand{\LLR}{\text{LLR}}

\newcommand{\comment}[1]
{\par {\small \bfseries \color{blue} #1 \par}} 

\newcommand{\bin}[2]{
    \left (
        \begin{array}{@{}c@{}}
        #1 \\ #2
        \end{array}
    \right )
}

\maketitle

\begin{abstract}
Despite of the known gap from the Shannon's capacity, several standards are still employing QAM or star shape constellations, mainly due to the existing low complexity detectors. In this paper, we investigate the low complexity detection for a family of QAM isomorphic constellations. These constellations are known to perform very close to the peak-power limited capacity, outperforming the DVB-S2X standard constellations. The proposed strategy is to first remap the received signals to the QAM constellation using the existing isomorphism and then break the log likelihood ratio computations to two one dimensional PAM constellations. Gains larger than 0.6 dB with respect to QAM can be obtained over the peak power limited channels without any increase in detection complexity. Our scheme also provides a systematic way to design constellations with low complexity one dimensional detectors. Several open problems are discussed at the end of the paper. 
\end{abstract}


\IEEEpeerreviewmaketitle

\section{Introduction}
\label{sec:INTR}

High order constellations with up to 256 signals have been adopted in new satellite communication standards such as digital video broadcasting (DVB-S2X) \cite{DVBS}, in view of the growing demand for spectral efficiency. Given the fact that the traffic demand for satellite broadband is expected to grow six-fold by 2020 \cite{BATS}, and the continuous need for higher data rates in satellite communications, even larger constellations may be needed in near future.

The receiver's computational complexity is one of the main concerns in any high throughput communication systems. On one hand, to increase the spectral efficiency one needs to employ high order constellations. On the other hand, the detection computational complexity grows as a function of constellation size and dimension. Indeed, despite the known gap from the Shannon capacity, quadrature amplitude modulation (QAM) constellations are still widely used in some standards (see for example \cite{DVB-T2}) due to the existing low complexity and simple detector. 
In this paper, we focus on bit-interleaved coding and modulation (BICM) scheme 
\cite{BICM}. In such systems, the log-likelihood ratio (LLR) computations are the receiver's computational bottleneck, specially for large constellations. 

The importance of receiver's complexity becomes more evident considering that constellations as large as 4096-QAM are already adopted in some standards \cite{DVB-C2} and they may be considered in future satellite systems too. In general, to compute the LLRs one needs to calculate the distance of the received signal from all the constellation points. Without any refinements, the complexity is of the order of the constellation size $M$, i.e., $\mathcal{O}(M)$. By projecting the $M$-QAM into two independent $\sqrt{M}$-PAM constellations (I and Q components), the complexity can be reduced to $\mathcal{O}(\sqrt{M})$ without any compensation in the performance over the additive white Gaussian noise (AWGN) channel \cite{QAMO}.

The QAM constellations perform rather poorly over the peak power limited channels due to their shapes and poor peak to average power ratio (PAPR). Peak power limited constraint is a common assumption in satellite communication systems with on-board high power amplifiers (HPA). The capacity achieving distribution under the peak power constraint is calculated in \cite{Shamai95}. Even though the optimal distribution for a finite set is not known theoretically, several studies indicate that amplitude and phase-shift keying (APSK) modulations perform very close to the capacity \cite{Tanaka}-\cite{GC2012}.

APSK constellations, in general, do not allow for low complexity detectors without a substantial performance degradation. Several authors have proposed ad hoc solutions for a given constellation or have focused on a specific subfamily of APSKs. For example, in \cite{KIM1} and \cite{KIM2} some techniques to reduce LLRs computations for DVB-S2X constellations are proposed. However, these techniques are specific to the constellations and the given bit to symbol mapping. Another important subfamily is the so called product-APSK (or star shape APSK), for which low complexity detection (LCD) has been studied by several authors \cite{TVehicular13}-\cite{softdemap2}. The main drawback of the product-APSK constellations is their inferior performance with respect to the optimized APSK constellations, mainly because number of points are fixed and constant over each ring. It is important to notice that the existence of a Gray mapping plays an important role in all these techniques.    

In this paper we investigate an alternative design strategy which allows for systematic low complexity detection. The proposed scheme is versatile, in the sense that the receiver may decide to choose the more complex ML detector with a better performance or the sub-optimal LCD. 
This versatility becomes important specially when on-board manipulation of the coded signal is needed.
The main idea is based on isomorphic maps between unit squares and unit disks in $\mathbb{R}^2$. We focus on a family of constellations obtained as the image of $M$-QAM under the radial map introduced in \cite{QCI_1}. We refer to these constellations as $M$-QCI (QAM to circular isomorphic). These constellations have a non-uniform APSK shape and they perform better than the optimized APSK constellations in DVB-S2X standard \cite{DVBS}. We kindly refer the readers to  \cite{QCI_1} and \cite{QCI_2} for more details on the performance of QCI constellations.

The existing isomorphic map between $M$-QAM and $M$-QCI allows to map the received signals back to the QAM constellation before computing the LLRs. By doing this operation, the noise distribution is no longer Gaussian. However, our results indicate that assuming a Gaussian distribution and breaking the LLR computations into two $\sqrt{M}$-PAM (I and Q components) will only cause a small loss in performance. Some further refinements of our method together with several open problems are discussed at the end of the paper.  

The rest of this paper is organized as follows. In Section \ref{sec:basics} we describe the notations and present our system model. We also briefly describe the QCI constructions. In Section~\ref{sec:Results} we present the simulation results and compare bit error rate (BER) of the proposed scheme with those of DVB-S2X constellations. We discuss some of the ongoing and future research directions in Section \ref{sec:future}. Finally, we conclude the paper in Section \ref{sec:conclude}.
\section{Basic concepts and notations} \label{sec:basics}

A constellation $\chi$ is a finite subset of the $k$-dimensional Euclidean space, i.e., $\chi \subset \mathbb{R}^k$. In this paper, we focus on constellations with dimensions 1 and 2. The elements of $\chi$ are referred to as symbols or signals. The symbols are associated to the bits at the input of the modulator through a one-to-one labelling $\mu: \chi \rightarrow \{ 0,1 \}^m$. For any given symbol $x$,  we denote by $\mu^{i}(x)$ the value of the $i^{th}$ bit of the label associated to it. $\chi_{b}^{i}$ denotes the 
subset of signals for which the labelling has value $b$ in
$i^{th}$ position, i.e., $\chi_{b}^{i} = \{ x \in 
\chi | \mu^{i}(x) = b \}$ ).

We denote respectively by $d(.,.)$ and $d_H(.,.)$ the Euclidean distance between two signals and the Hamming distance between two 
binary sequences.
A labeling for $\chi$ is called a Gray labeling if for 
any two signal $x_i,x_j \in \chi$ we have
$d_H(\mu(x_i),\mu(x_j)) = 1$ if  $ d(x_i,x_j) \leq  
d(x_i,x_k)$, for all $x_k \in \chi$.

\subsection{LLR computations and detection complexity}
We consider an additive white Gaussian noise (AWGN) channel with the received signal $y=x+n$, where $n$ has a Gaussian distribution with zero mean and variance $N_0/2$ per dimension.

In BICM systems, at the receiver, we need to calculate the LLRs for each coded bit based on the received signal $y$. The LLRs are then passed to the decoder as the soft information. For each bit $i$ the LLR$_i$ can be expressed as:
\begin{eqnarray}\label{eq:LLR}
    \LLR_i &=&  \log \left(\frac{\sum_{x \in \chi^{i}_{0}}p(y|x)}{\sum_{x \in
\chi^{i}_{1}}p(y|x)}\right).
\end{eqnarray}
Notice that $\chi^{i}_{0}$ and $\chi^{i}_{1}$ are complementary sets, and therefore we need to calculate $p(y|x)$ for all given symbols $x \in \chi$. A rigorous estimation of the computational complexity of $\LLR_i$'s would need to take into account the total number of needed operation and their type. For simplicity, in this paper, we assume that the main source of the complexity at the receiver is to compute the $p(y|x)$, or $d(x,y)$. Therefore, the computational complexity of $\LLR_i$ is roughly of order $\mathcal{O}(M)$, where $M=2^m$ is the cardinality of the constellation set. For more detailed analysis of the computational complexity of $\LLR_i$ we kindly refer the readers to \cite{TVehicular13} and references within.

One of the main characteristics of $M$-QAM modulation scheme is that the $\LLR$ computations can be broken into two independent I and Q components. By doing so, for each of the components, the LLRs are separately computed over a $\sqrt{M}$-PAM constellation without any performance degradation. Therefore, the computational complexity of $\LLR_i$ for a $M$-QAM constellation can be reduced to $\mathcal{O}(\sqrt{M})$. This property allows to use very high order QAM constellations without overloading the receiver's computational complexity \cite{DVB-C2}. In the rest of this paper we use the term low complexity detection (LCD) to indicate that the $\LLR$ computations are of order $\mathcal{O}(\sqrt{M})$.

\subsection{QAM isomorphic constellations}
In \cite{QCI_1} a family of circular constellations has been introduced. 
These constellations, named QAM to circular isomorphic (QCI), can be described as the QAM image under the radial isomorphosim. This isomorphism maps the concentric squares into the concentric circles. The same map can be used to project the unit square onto a Disc. In more details, let $S = \{ (u,v) \in \mathbb{R}^2 \;| \;|u| \leq 1 ,|v| \leq 1 \}$ and $C = \{ (u,v) \in \mathbb{R}^2 \;| \; u^2+v^2 \leq 2 \}$ denote the unit square and the disc with radios $\sqrt{2}$, respectively. Then the radial map $f:S \rightarrow C$  can be written analytically as below:
\[
 f(u,v) =
  \begin{cases} 
      \hfill \frac{\sqrt{2} \max (|u|,|v|)}{\sqrt{u^2+v^2}} (u,v)     \hfill & \text{ if $(u,v) \neq (0,0)$ } \\
      \hfill (0,0) \hfill & \text{ if $(u,v) = (0,0)$} \\
  \end{cases}
\]
We denote the inverse map by $f^{-1} = 1/f$ for $(u,v)\neq(0,0)$ and $f^{-1}(0,0)=(0,0)$. It is important to notice that the binary Gray labelling of QAM constellation will be preserved under $f$, and the resulting QCI constellation has also a Gray labelling. Using this isomorphism, for each $M$-QAM constellation, a unique $M$-QCI constellation can be constructed.

\subsection{Peak signal to average noise power ratio}
Nonlinear characteristics of the satellite channel largely
depend on the HPA used at the satellite transponder and
operating close to the saturation point. In the following we assume a hard-limiter instead of the real AM/AM amplification curves where the AM/AM curve is linear up to a saturation point where it becomes a constant. In principle, we are interested in the case that the HPA
is acting very near to its saturation point. In this case, the parameter that characterize the satellite communication system is Peak signal power to average noise power ratio (PSNR) defined as:

$$  \PSNR \triangleq \frac{1}{N_0}\geq \SNR \triangleq
\frac{1}{M}\sum_{x \in\chi} \frac{|x|^2}{N_0}.
$$
Assuming a memoryless ideal non-linearity model for the HPA and ignoring the effect of filters, the PSNR coincides with $P_{sat}/N_0$: 
$P_{sat}/N_0$:
$$
\SNR = \frac{E_s}{N_0} = \frac{P_{\rm sat}}{N_0}\cdot\frac{E_s}{P_{\rm sat}} = \PSNR - OBO \;\; {\text {[dB]}}.
$$
For a detailed derivation we kindly refer the readers to \cite{GC2012}. In this paper we plot the BER simulation results as a function of PSNR.  the average power constraint over the signal space is substituted by the
maximum power constraint.

\subsection{Proposed detection scheme}
As we will see in section~\ref{sec:Results}, the $M$-QCI constellations perform slightly better than the constellations adopted in DVB-S2X standard employing the ML detector. In this paper, we propose an alternative LCD scheme for QCI constellations. The main idea is to remap the received signals back to the QAM constellation by $f^{-1}$ before computing the LLRs. By doing so, one can break the LLR computations into two $\sqrt{M}$-PAM (I and Q components). 

\begin{figure}[tbh]
\vspace*{-1.1in}
\centerline{\includegraphics[scale=.5]{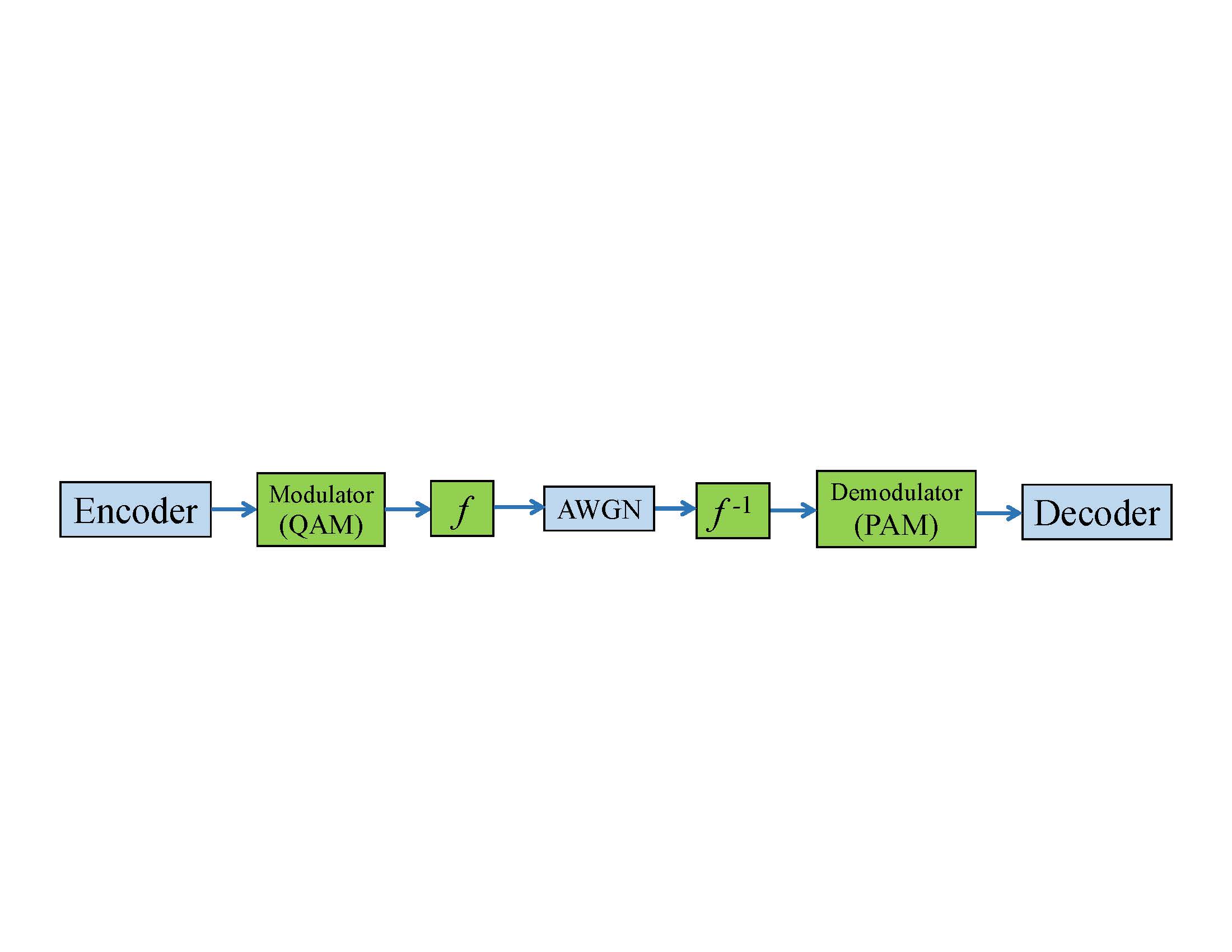}}
\vspace*{-1.0in}
\caption{Block diagram for the proposed modulation/demodulation scheme.}
\label{fig:blockdiagram}
\end{figure}

The block diagram for the proposed LCD modulation scheme for QCI constellations is presented in Fig.~\ref{fig:blockdiagram}. We first map the coded bits into the $M$-QAM before passing it through the $f$ radial map. Notice that $f$ does not change the peak power of the QAM constellation. At the receiver, the received signal is passed through the $f^{-1}$ block before being decomposed into I and Q components. As we will see later, the noise at the output of the $f^{-1}$ is no longer Gaussian and the decomposition into I and Q is not lossless.

\section{Simulation Results}\label{sec:Results}
In this section we present the BERs of the proposed LCD scheme for $M$-QCI constellations with $M=16, 64$ and 256. We compare the results with  QAM and APSK constellations. 
In all the simulations the low-density parity-check code (LDCP) of DVB-S2X with rate 3/4 is used. 

\begin{figure}[tbh]
\vspace*{-0.3in}
\centerline{\includegraphics[scale=.5]{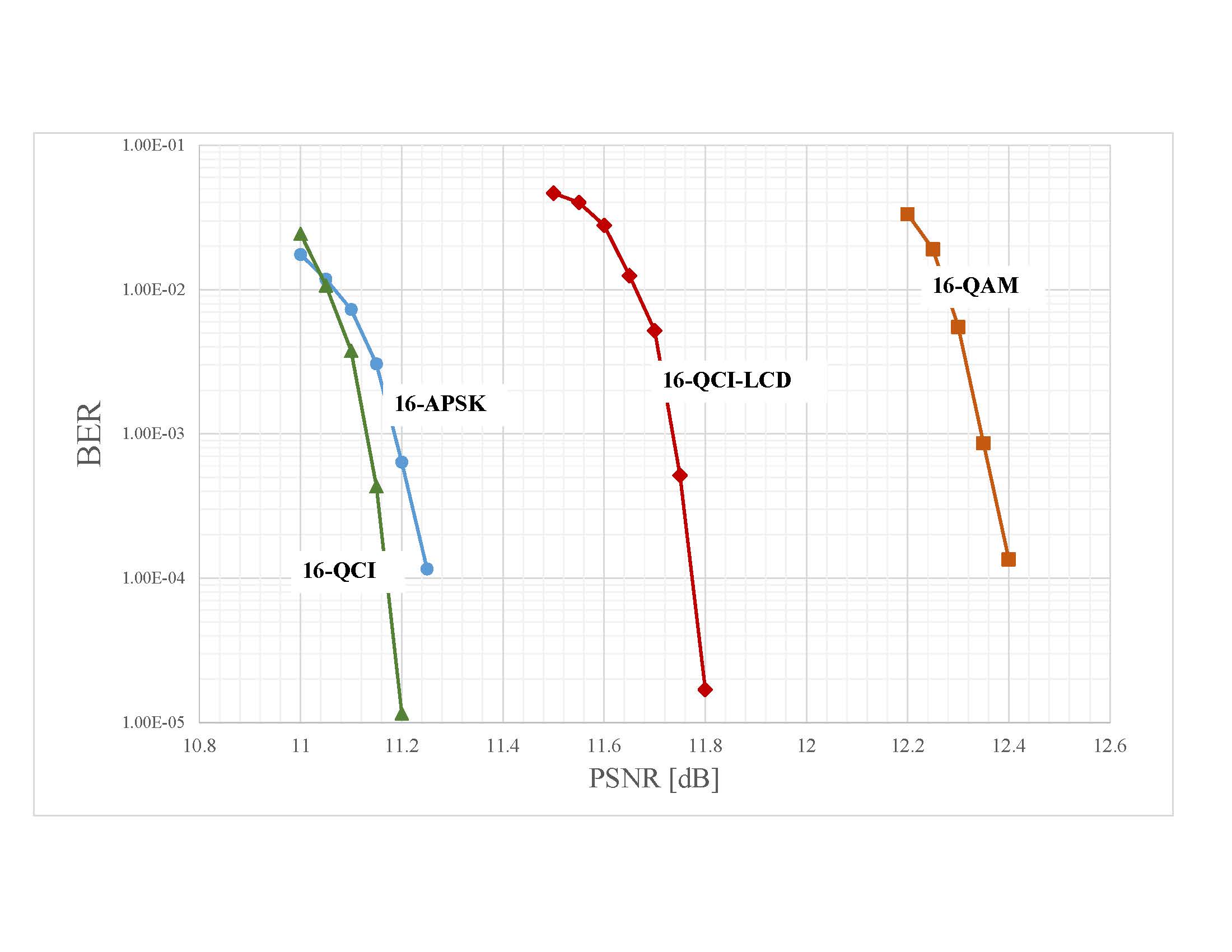}}
\vspace*{-0.3in}
\caption{BER result for 16-QCI constellations with LCD and comparison with 16-QAM and 16-APSK.}
\label{fig:16points}
\end{figure}

\begin{figure}[tbh]
\vspace*{-0.3in}
\centerline{\includegraphics[scale=.5]{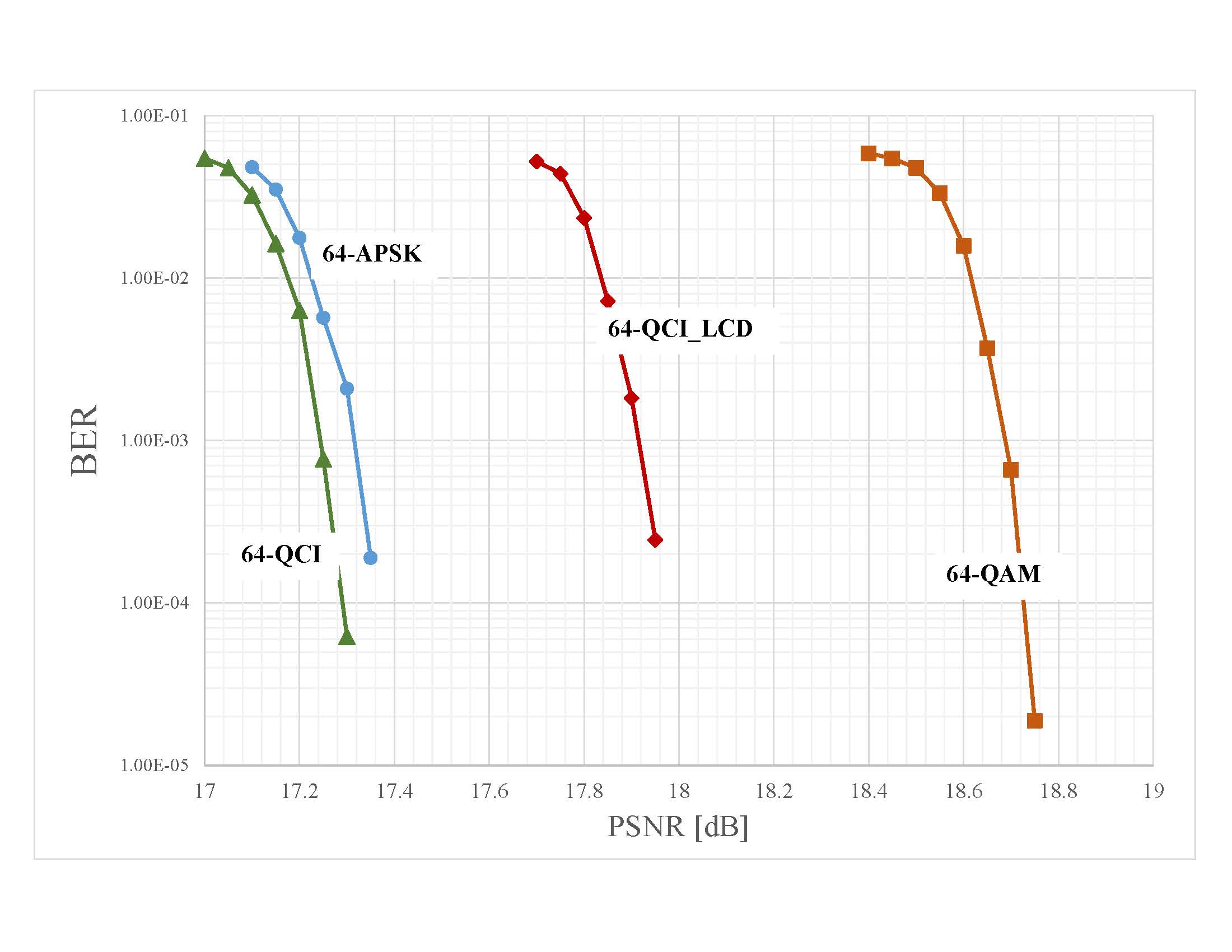}}
\vspace*{-0.2in}
\caption{BER result for 64-QCI constellations with LCD and comparison with 64-QAM and 64-APSK.}
\label{fig:64points}
\end{figure}

\begin{figure}[tbh]
\vspace*{-0.3in}
\centerline{\includegraphics[scale=.5]{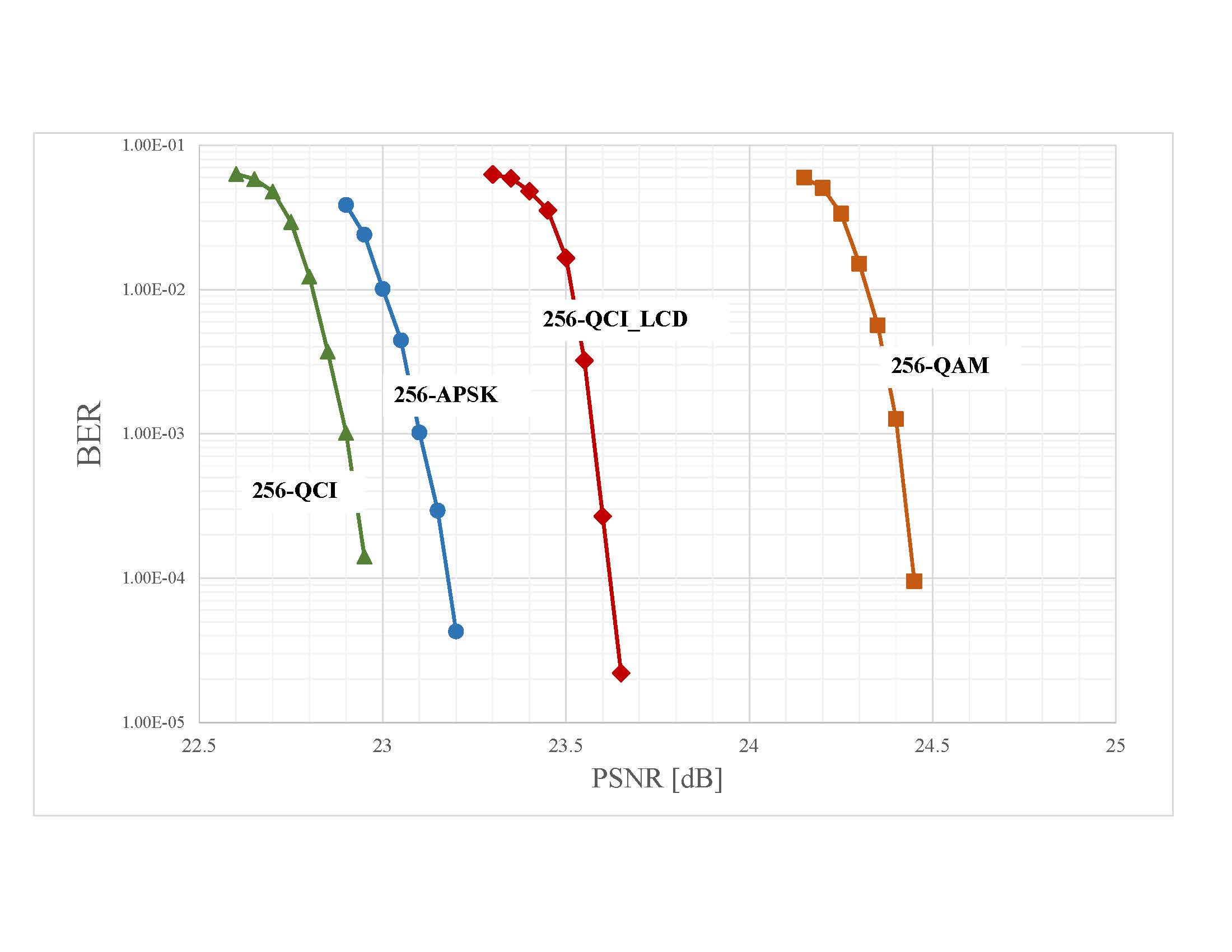}}
\vspace*{-0.3in}
\caption{BER result for 256-QCI constellations with LCD and comparison with 256-QAM and 256-APSK.}
\label{fig:256points}
\end{figure}

In Fig.~\ref{fig:16points} the results for constellations with 16 points are presented. The 16-QCI performs slightly better than the 16-APSK constellation with ML detector. Moreover, BER curve of the 16-QCI has a steeper slope. The loss of using the LCD scheme of Fig. \ref{fig:blockdiagram} with respect to the ML detector is around 0.6 dB. However, LCD scheme still performs around 0.6 dB better than the 16-QAM. It is important to notice that this gain is obtained without increasing the detector's complexity compared to QAM. 
Same simulations are done for constellations of orders 64 and 256. The results are shown respectively in Figs. \ref{fig:64points} and \ref{fig:256points}. As it can be seen, the gain with respect to 64-QAM and 256-QAM are slightly larger in these cases, while the loss with respect to ML detection remains the same. It is also important to notice that 256-QCI constellation outperforms the 256-APSK constellation of DVB-S2X by 0.2 dB.

\section{Discussions for further improvements}\label{sec:future}
\begin{figure}[tbh]
\vspace*{-0.19in}
\centerline{\includegraphics[scale=.35]{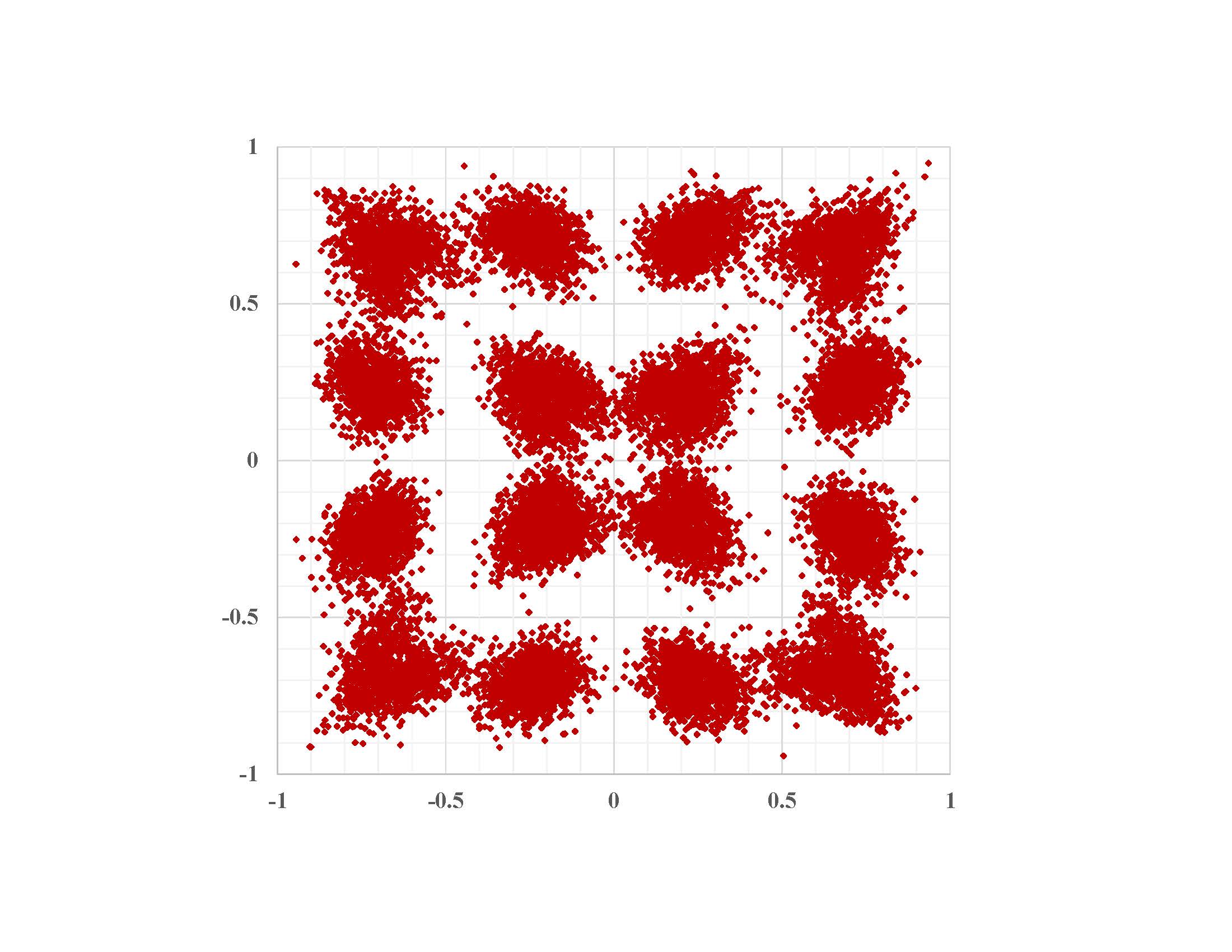}}
\vspace*{-0.25in}
\caption{Scatter plot at the output of the $f^{-1}$ block.}
\label{fig:scatterplot}
\end{figure}

Proposed LCD scheme can be further improved without increasing the complexity. Here we try to explain briefly the main assumptions and approximations which are involved in this scheme. First of all, we should notice that LLRs are computed based on the signal $f^{-1}(y) = f^{-1}(x+n)$ where $x$ is chosen from the QCI constellation. It is easy to show that the radial map does not satisfies the Cauchy's functional equation, i.e., $f^{-1}(x+n) \neq f^{-1}(x)+f^{-1}(n)$. Therefore, in general, $f^{-1}(y)$ cannot be considered as a QAM signal plus noise. Moreover,  $f^{-1}(n)$ is not any more Gaussian, and in principle, both the mean and variance of the noise is changed at the output of the $f^{-1}$ block in Fig. \ref{fig:blockdiagram}. In other words, the BER results presented in the previous section are obtained by a mismatched demodulator. To visualize this, the scatter plot at the output of $f^{-1}$ block is shown in Fig. \ref{fig:scatterplot}. From this figure, it can also be observed that the noise distribution may not be written as the Cartesian product of two independent distributions. 

\begin{figure}[tbh]
\vspace*{-0.19in}
\centerline{\includegraphics[scale=.35]{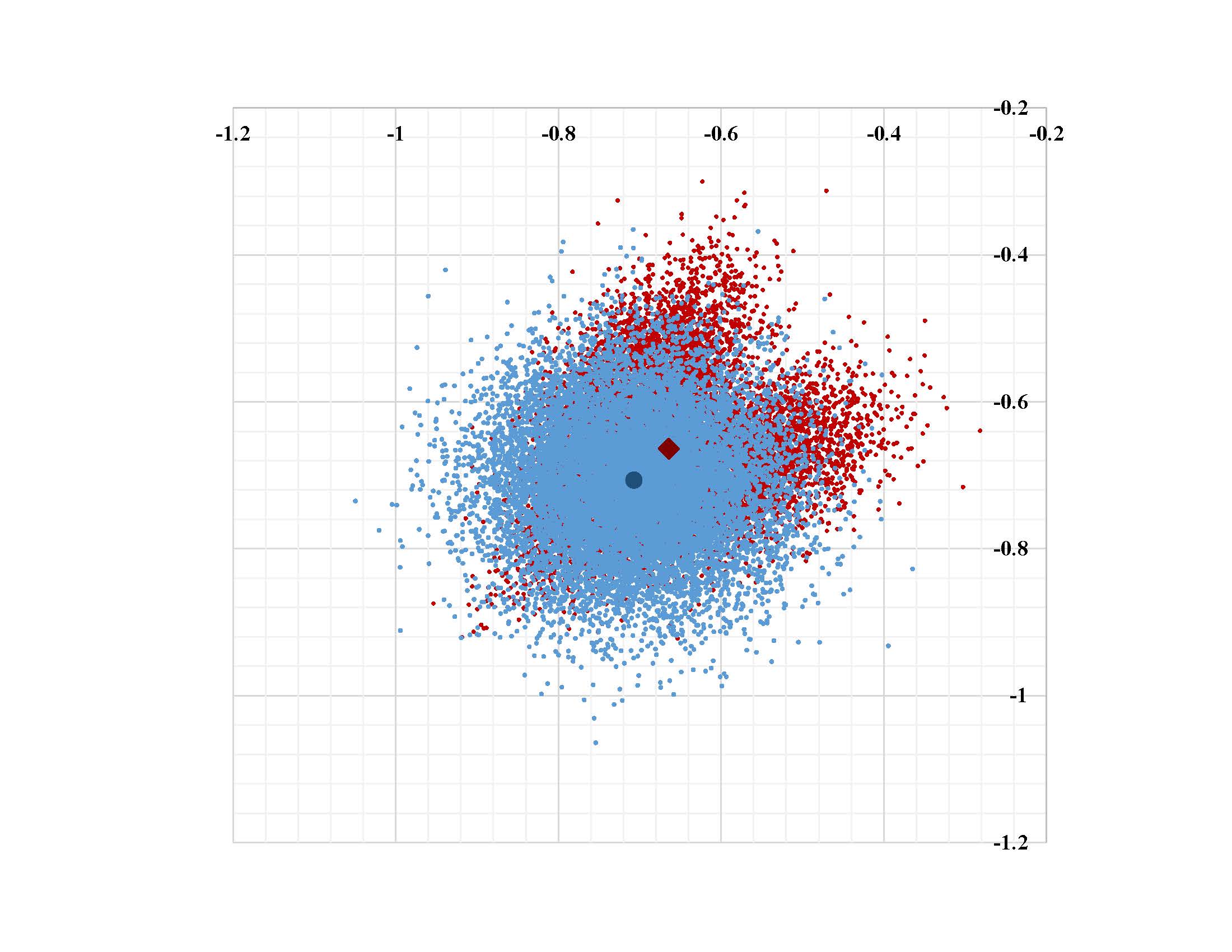}}
\vspace*{-0.1in}
\caption{Berry center of the QAM scatter plot does not match with that of the LCD QCI scheme.}
\label{fig:average}
\end{figure}

In order to highlight better the fact that reference QAM constellation at the modulator is different from the one at the receiver in the proposed scheme, in Fig. \ref{fig:average} we plot the berry centers of the respective clouds for a corner point of the QAM constellations. The red cloud corresponds to that of the $f^{-1}$ and the blue cloud corresponds to the QAM signal plus the Gaussian noise. We show the berry centers of each cloud by a red square and a blue circle respectively. 
It is an interesting problem to calculate analytically the $k$-order statistics of $f^{-1}(x+n)$ for each signal $x$. However, this may be a difficult problem, and the statistics may also depend on the signal $x$. A simpler problem is to find $\alpha$ and $\beta$ such that the average and variance of $\alpha f^{-1}+\beta$ provides the best approximation for the channel noise. Some preliminary results indicate that gains around 0.15 dB can be obtained by estimating $\alpha$ alone. Research in this direction is on going. 

\begin{figure}[tbh]
\vspace*{-0.3in}
\centerline{\includegraphics[scale=.5]{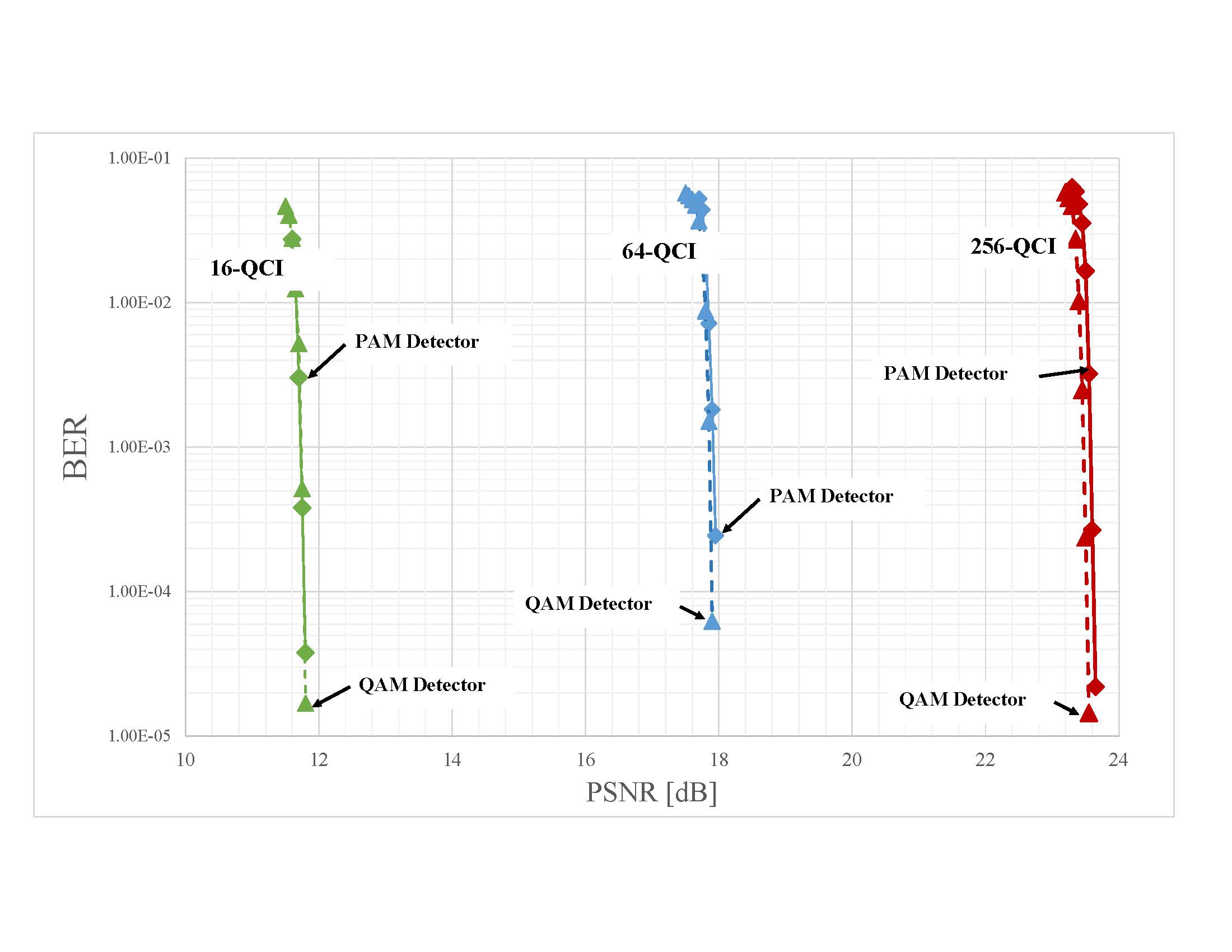}}
\vspace*{-0.26in}
\caption{The loss due to I and Q decomposition at the demodulator $M$-QCI constellations.}
\label{fig:IandQ}
\end{figure}

Another interesting problem is to investigate how much of the loss in LCD scheme is due to non-optimal I and Q decomposition. To answer this question, we have calculated the optimal LLRs for $f^{-1}(x+n)$ signal, without decomposing it into I and Q component in our scheme. The results are presented in Fig. \ref{fig:IandQ} for various constellation orders. The dashed lines correspond to the optimal LLR calculation in 2D. As it can be observed, the loss due to I and Q decomposition is almost negligible (smaller than 0.1 dB in all cases), indicating that the main source of performance loss is due to the mismatched demodulation. This fact is actually quite surprising and needs to be further studied in details, as it may indeed suggest new methodology for inventing new LCD schemes using other types of isomorphisms. 

\section{Conclusions}
\label{sec:conclude} 
In this paper we have investigated a low complexity detection scheme for QCI constellations by using the inverse radial map before LLR computations. We show that gains up to 0.8 dB with respect to the QAM constellations can be obtained under the peak power constraint without increasing the detectors complexity. On the other hand, the loss due to non-optimal detection is around 0.6 dB. The main reason for the loss is shown to be the mismatched demodulation at the receiver and the loss caused by decomposition into I and Q components is negligible. Our results indicate that it is possible to improve the performance of QAM constellations over the peak power limited channel by shaping the constellation without increasing the detector complexity. Our preliminary results indicate that by estimating the noise parameters at the demodulator, larger gains can be obtained. 
\balance
\section*{Acknowledgment}
The author would like to thank Guido Montorsi and Giulio Colavolpe for several inspiring discussions. The author is partially supported by the Luxembourg National Research Fund under CORE Junior project: C16/IS/11332341 (ESSTIMS). This research is partially supported by the European Space Agency (ESA) under the TRP contract: ITT A0/1-8332/15/NL/FE (OPTIMUS). The views of the author do no reflect the views of ESA.

\bibliographystyle{IEEEtran}

%
\end{document}